\documentclass[iop]{emulateapj}

\def\Msun{M_\odot}
\def\microas{\mu{\rm as}}

\begin{document}

\title{The PHASES Differential Astrometry Data Archive IV:  The Triple Star Systems 63 Gem A and HR 2896}

\author{Matthew W.~Muterspaugh\altaffilmark{1, 2}, 
Francis~C.~Fekel\altaffilmark{2}, 
Benjamin F.~Lane\altaffilmark{3}, 
William I. Hartkopf\altaffilmark{4}, 
S.~R.~Kulkarni\altaffilmark{5}, 
Maciej Konacki\altaffilmark{6, 7}, Bernard F.~Burke\altaffilmark{8}, 
M.~M.~Colavita\altaffilmark{9}, 
M.~Shao\altaffilmark{9}, 
M.~Williamson\altaffilmark{2}}
\altaffiltext{1}{Department of Mathematics and Physics, College of Arts and 
Sciences, Tennessee State University, Boswell Science Hall, Nashville, TN 
37209 }
\altaffiltext{2}{Tennessee State University, Center of Excellence in 
Information Systems, 3500 John A. Merritt Blvd., Box No.~9501, Nashville, TN 
37209-1561}
\altaffiltext{3}{Draper Laboratory,  555 Technology Square, Cambridge, MA 
02139-3563}
\altaffiltext{4}{U.S.~Naval Observatory, 3450 Massachusetts Avenue, NW, Washington, DC, 20392-5420}
\altaffiltext{5}{Division of Physics, Mathematics and Astronomy, 105-24, 
California Institute of Technology, Pasadena, CA 91125}
\altaffiltext{6}{Nicolaus Copernicus Astronomical Center, Polish Academy of 
Sciences, Rabianska 8, 87-100 Torun, Poland}
\altaffiltext{7}{Astronomical Observatory, Adam Mickiewicz University, 
ul.~Sloneczna 36, 60-286 Poznan, Poland}
\altaffiltext{8}{MIT Kavli Institute for Astrophysics and Space Research, 
MIT Department of Physics, 70 Vassar Street, Cambridge, MA 02139}
\altaffiltext{9}{Jet Propulsion Laboratory, California Institute of 
Technology, 4800 Oak Grove Dr., Pasadena, CA 91109}

\email{matthew1@coe.tsuniv.edu, wih@usno.navy.mil, blane@draper.com, maciej@ncac.torun.pl}

\begin{abstract}
Differential astrometry measurements from the Palomar High-precision 
Astrometric Search for Exoplanet Systems (PHASES) are used to constrain the 
astrometric orbit of the previously known 
$\lesssim 2$ day subsystem in the triple system 63 Gem A and have detected a 
previously unknown 2 year Keplerian wobble superimposed on the visual 
orbit of the much longer period (213 years) binary system HR 2896.
63 Gem A was already known to be triple 
from spectroscopic work, and absorption 
lines from all 3 stars can be identified and their individual Doppler shifts 
measured; new velocities for all three components are presented to aid in 
constraining the orbit and measuring the 
stellar masses.  In fact, 63 Gem itself 
is a sextuple system:  the hierarchical 
triple (Aa1-Aa2)-Ab (in which Aa1 and Aa2 orbit 
each other with a rapid period just under 2 days, and Ab orbits these every 2 
years), plus three distant common proper motion companions.  
The very small astrometric perturbation caused by the inner pair in 63 Gem A 
stretches the limits of current astrometric capabilities, but PHASES 
observations are able to constrain the orientation of the orbit.  
The two bright stars comprising the HR 2896 long period (213 year) system have 
a combined spectral type of K0III and the newly detected object's mass estimate 
places it in the regime of being a M dwarf.  The motion of 
the stars are slow enough that their spectral features are always blended, 
preventing Doppler studies.  The PHASES measurements and radial velocities 
(when available) have been combined with lower precision single-aperture 
measurements covering a much longer timeframe (from eyepiece measurements, 
speckle interferometry, and adaptive optics) to improve the characterization of 
the long period orbits in both binaries.  The visual orbits of the short and 
long period systems are presented for both systems, and used to calculate two 
possible values of the mutual inclinations between inner and outer orbits of 
$152 \pm 12$ degrees or a less likely value of $31   \pm 11 $ degrees for 
63 Gem A and $10.2 \pm 2.4$ degrees or $171.2 \pm 2.8$ degrees for HR 2896.  
The first is not coplanar, whereas the second is either nearly coplanar or 
anti-coplanar.
\end{abstract}

\keywords{binaries:close -- binaries:visual -- astrometry}

\section{Introduction}

There are two major motivating factors for studying systems with three or 
more stars.  First is the determination of the fundamental properties of the 
stars themselves.  Their masses, luminosities, and radii can be derived if 
high resolution imaging and radial velocity (RV) orbits are combined.  
Through their physical association, one can assume all stars in the system 
have identical ages, primordial elemental abundances, and similar histories.  
The only variations that can explain observed differences are the masses of 
the stars themselves.  Binary stars have been important for establishing 
constraints on modelling stellar structure and evolution.  These constraints 
become much more strict as the number of stars in a system grows---the science 
gained scales much faster than linear with the number of system components.

Second is the complex dynamics in systems with more than two components.  The 
planets of our solar system are found in a flat disk.  There is great 
interest in determining whether multiple star systems are equally coplanar 
because this tells about the formation environments of stars 
\citep{Sterzik2002} and their subsequent evolutions \citep{Fabrycky2007}.  
From the 6 triples and 2 quadruples studied so far, it is 
clear that the distribution is neither always coplanar nor is it consistent 
with randomly oriented subsystems.

To determine unambiguously a system's coplanarity, one must have both 
visual and RV orbital solutions for pairs of interest.  The reason that mutual 
inclination measurements have been rare is because of the observational 
challenges these systems present.  RV signals are largest for compact pairs 
of stars, whereas imaging prefers wider pairs.  The ``wide'' pair must have 
large enough RV amplitude for a signal to be detected, thus its physical (and 
apparent) separation is as small as the two-component binaries that are 
already challenging for visual studies.  The ``narrow'' pair is even smaller.  
This paper presents the geometries of two new triple star systems:  63 Gem A 
and HR 2896.

This paper is the fourth in a series, analyzing the final results of the 
Palomar High-precision Astrometric Search for Exoplanet Systems (PHASES)
project after its completion in late 2008.  The first paper describes the 
observing method, sources of measurement uncertainties, limits of observing 
precisions, derives empirical scaling rules to account for noise sources 
beyond those predicted by the standard reduction algorithms, and presents the 
full catalog of astrometric measurements from PHASES \citep{Mute2010A}.  The 
second paper combines PHASES astrometry with astrometric measurements made by 
other methods as well as radial velocity observations (when available) to 
determine orbital solutions to the binaries' Keplerian motions, determining 
physical properties such as component masses and system distance when 
possible \citep{Mute2010B}.  The third paper 
presents limits on the existence of substellar tertiary companions, orbiting 
either the primary or secondary stars in those systems, that are found to be 
consistent with being simple binaries \citep{Mute2010C}.  The current paper 
presents three-component orbital solutions to a known triple star system 
(63 Gem A $=$ HD 58728) and a newly discovered triple system 
(HR 2896 $=$ HD 60318).  Finally, paper five presents candidate substellar 
companions to PHASES binaries as detected by astrometry \citep{Mute2010E}.

63 Geminorum (HR 2846, HD 58728, ADS 6089, HIP 36238) is a hierarchical 
multiple system of 6 components.  Its components have been studied in a 
variety of ways, including as spectroscopic, visual, common proper motion, 
and lunar occultation binaries.  Component A, with composite spectral 
type F5 IV-V, is a sub-arcsecond hierarchical triple system (the components 
are Aa1, Aa2, and Ab), with additional faint common proper motion companions B, 
C, and D at distances of 43, 146, and 4 arcseconds, respectively.  The 
proper motions of A, B, and D are all similar and likely to be a true multiple 
system, but C is probably just an optical companion.  This paper 
reports the three dimensional orbits of the triple subsystem; Components B, 
C, and D are not considered in the rest of the current study.  The three 
components of subsystem A are organized as follows.  A itself is split into 
two objects, designated Aa and Ab, which orbit each other with a period of 
$\sim 760$ days.  One of these, Aa, is itself a binary with components 
designated Aa1 and Aa2, which orbit each other with period $\sim 1.9$ days.  
The wider pairing, Aa-Ab, was resolved with speckle interferometry 
\citep{McA1983} and has been studied almost exclusively by speckle 
differential astrometry, with the occasional lunar occultation measurement.  
The short period system has not been previously resolved, but has been 
monitored with RV.  The high precision ($\sim 35 \microas$) differential 
astrometry technique developed by \cite{LaneMute2004a} has been used to 
measure the position vector between the Aa1-Aa2 center-of-light (COL) and the 
location of star Ab as part of PHASES \citep[][]{Mute06Limits}.  These 
measurements provide a detailed visual orbit of the Aa-Ab 760 day system, and 
also detect the 1.9 day COL wobble of the Aa1-Aa2 pair; 
these orbits are presented here.  By 
combining astrometry and RV observations in a simultaneous fit, the mutual 
inclination between these pairs is measured, though with some ambiguity 
depending on whether Aa1 or Aa2 is more luminous at near-infrared K-band 
wavelengths (2.2 ${\rm \mu m}$).  Because Aa2 is less massive than Aa1, it is 
more likely that the solution favoring it being less luminous is correct.

HR 2896 (HD 60318, HIP 36896, WDS 07351$+$3058) was discovered to be a binary 
system with subarcsecond separation in 1842 by \cite{Stt1878}, 
though this result was not published until much later.  The first published 
measurement of its binarity was reported by \cite{Mad1844}.  There is a faint 
C component at 82 arcseconds, though it does not share a common proper motion 
and is probably not physical.  Since the subarcsecond system was discovered, a 
number of visual observers and speckle interferometry teams have monitored the 
binary motion to map its very long period orbit.  The latest program to monitor 
this pair, at yet higher astrometric precision, is PHASES, using 
long baseline interferometry and real-time phase-referencing techniques to 
measure binary separations with $\sim 35 \, {\rm \mu as}$ repeatability.  
Model fitting the observed motion with a single 
Keplerian model provides unsatisfactory agreement with these higher precision 
observations, prompting a search for additional components to the system.  
A visual inspection of the residuals shows an obvious trend with a 
two year period.  

Astrometric measurements were made at the Palomar Testbed Interferometer 
\citep[PTI;][]{col99}.  PTI was 
located on Palomar Mountain near San Diego, CA. It was developed by the Jet 
Propulsion Laboratory, California Institute of Technology for NASA, as a 
testbed for interferometric techniques applicable to the Keck Interferometer 
and other missions such as the Space Interferometry Mission (SIM).  It 
operated in the J ($1.2 \mu{\rm m}$), H ($1.6 \mu{\rm m}$), and K 
($2.2 \mu{\rm m}$) bands, and combined starlight from two out of three 
available 40-cm apertures.  The apertures formed a triangle with one 110 and 
two 87 meter baselines.  PHASES observations began in 2002 continued through 
November 2008 when PTI ceased routine operations.

\section{Astrometric Measurements}

\subsection{PHASES Measurements}

Twenty-one differential astrometry measurements of 63 Gem A spanning 1495 days 
(nearly 2 orbits of the wide binary system) and 13 measurements of HR 2896 
spanning 739 days were obtained with 
the procedure described by \cite{LaneMute2004a} and Paper I, with the use of 
the standard PHASES data reduction pipeline summarized in Paper I.  
No PHASES measurements are identified as significant outliers to the 
double Keplerian models, and all are used in orbit fitting.

Analysis of several PHASES binaries indicates that the formal measurement 
uncertainties evaluated by the standard data reduction pipeline underestimate 
the actual scatter in the measurements.  As described by Paper I, 
a process for modifying the measurement uncertainties was found that best 
reproduces the observed scatter about Keplerian models for several binary 
targets.  For measurements without the longitudinal dispersion compensator 
and/or automatic alignment system (as is applicable to all measurements of 
HR 2896), the solution is to increase the 
formal uncertainties along the error ellipse's major axis by a factor of 
1.3 and then add $140 \, {\rm \mu as}$ in quadrature, while for the minor 
axis the formal uncertainties are increased by a factor of 3.8 after which 
$35 \, {\rm \mu as}$ is added in quadrature.  When both the longitudinal 
dispersion compensator and automatic alignment system were operational (as was 
the case for about half of the 63 Gem A measurements), the same uncertainty 
corrections are applied except that the minor axis scaling factor is 1.0 
instead of 3.8.  The measurements and their 
associated formal and rescaled measurement uncertainties are listed in 
Paper I.  The average scaled measurement uncertainty 
is $61 \, {\rm \mu as}$ in the minor axis and $499 \, {\rm \mu as}$ in the 
major axis for 63 Gem A, and $40 \, {\rm \mu as}$ and $228 \, {\rm \mu as}$ for 
HR 2896, respectively.

\subsection{Non-PHASES Astrometry Measurements}

The wider pairing of 63 Gem A (Aa-Ab) was resolved for the first time with 
speckle interferometry by \cite{McA1983}.  This and other speckle 
interferometry measurements have been collected in the Washington Double Star 
Catalog 
\citep[henceforth WDS, ][and see references to the original works therein]{wdsCatalog, wdsCatalogUpdate}, 
and are presented in Table \ref{tab::speckleData}.  A few lunar occultation 
measurements have also been conducted, though they are not included 
in the present analysis because they have little impact on the final orbital 
solution.

Differential astrometry of HR 2896 has a long history.  Observations, spanning 
over 160 years, that used single-telescope techniques (including micrometry, 
speckle interferometry, and other interferometric techniques) have been 
cataloged in the WDS which includes 183 entries for HR 2896.  These 
measurements are included in the present analysis to better constrain the 
much longer period A-B orbit.

Measurement uncertainties were evaluated with the algorithm for assigning 
weights, described by \cite{hart01}.  Unit weight uncertainties were assigned 
separately for the separation and position 
angle measurements in such a way that the weighted root-mean-square 
scatter about a Keplerian model in each axis was unity.  Measurements 
identified as more than a $3$-$\sigma$ outlier in either separation or 
position angle were flagged as outliers, and the procedure was iterated until 
no outliers were found.  One of the 29 measurements of 63 Gem A was found to 
be an outlier, and the unit weight uncertainties are 14.7 mas in separation and 
$4.70^\circ$ in position angle.  The average uncertainties 
of the 28 measurements that were used are 11.4 mas and $3.7^\circ$, 
respectively, and the smallest values are 5.1 mas and $1.6^\circ$.  
Only two of the 183 measurements of HR 2896 were identified as outliers 
in this manner.  The unit weight uncertainties are 84.6 mas in separation and 
$2.98^\circ$ in position angle.  The average 
uncertainties of the 181 measurements 
being used are 86.4 mas and $3.04^\circ$; the 
smallest uncertainties are 16 mas and 
$0.57^\circ$.  The measurements, their assigned 
uncertainties, and weights are listed 
in Table \ref{tab::speckleData}.

\section{Spectroscopic Measurements}

\subsection{63 Gem A}

\cite{abt1976} reported 20 radial velocity (RV) measurements of 
component Aa1, and 17 of component Aa2 and determined the spectroscopic 
orbit of that subsystem.  The scatter in the residuals resulting from a 
double-Keplerian model fit to these data suggests that the measurement 
uncertainties may be underestimated by a factor of 2.21 for component Aa1 
and 3.83 for Aa2.  The formal uncertainties reported in that work have 
been increased by these factors in order that this data set can be used in 
combined fits with others.

Twenty-six additional velocity measurements of all three components, and 2 of 
Aa1 and Ab, have been collected at Kitt Peak with 
the Coud\'e Feed Telescope and spectrograph, spanning 8 years.  Spectral 
features of all three components are present in these high 
quality spectra.  
Tennessee State University's 2-m Automatic Spectroscopic Telescope 
\citep[AST, ][]{eatonWilliamsonAST} 
observed 63 Gem A 45 times over a 6.3 year period, spanning 
3 orbits of the longer period Aa-Ab system.  Spectral features of all three 
components are observed, and for each epoch the velocity of each star can 
be measured.
All these spectroscopic measurements are used in the present analysis and 
are listed in Table \ref{tab::rv_63_Gem} with the measurement uncertainties 
used during orbit fitting.

\subsection{HR 2896}

TSU's AST obtained spectra of HR 2896 on MJD 55267, 55272, 55276, 55286, 
55295, 55304, 55321, and 55329.  Only one set of absorption lines were 
observed, indicating blended lines and Doppler shifts too small 
to identify multiple components.

\section{Searching for the Short Subsystem Orbital Periods}

\begin{figure*}[!ht]
\plottwo{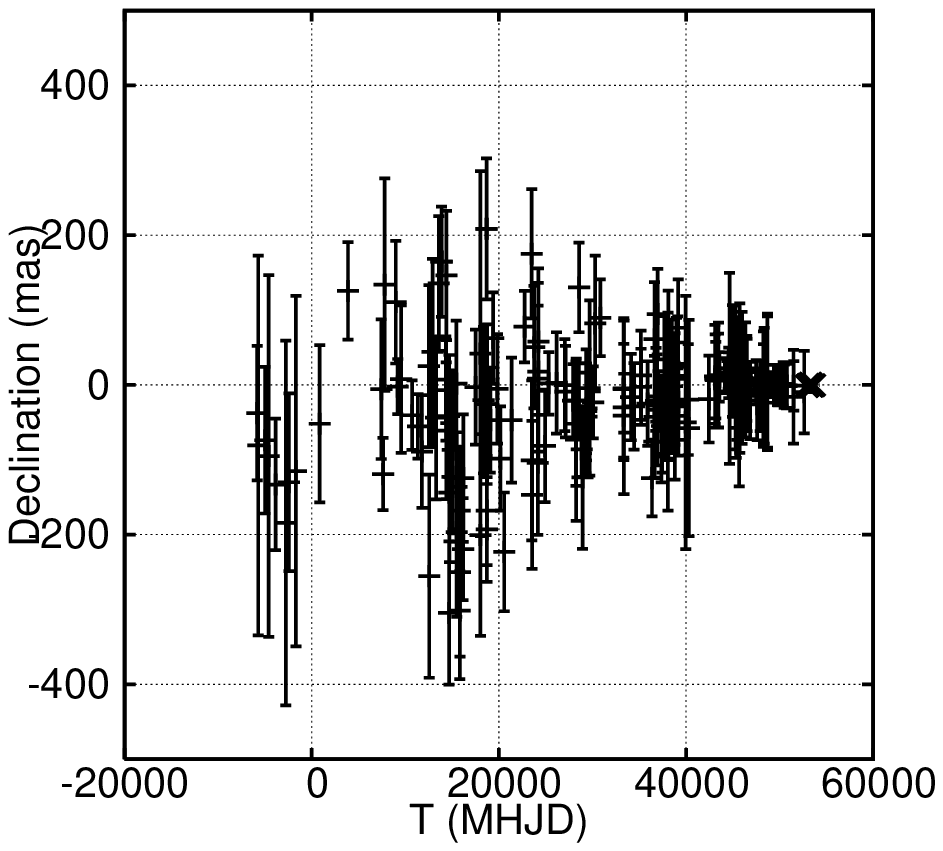}{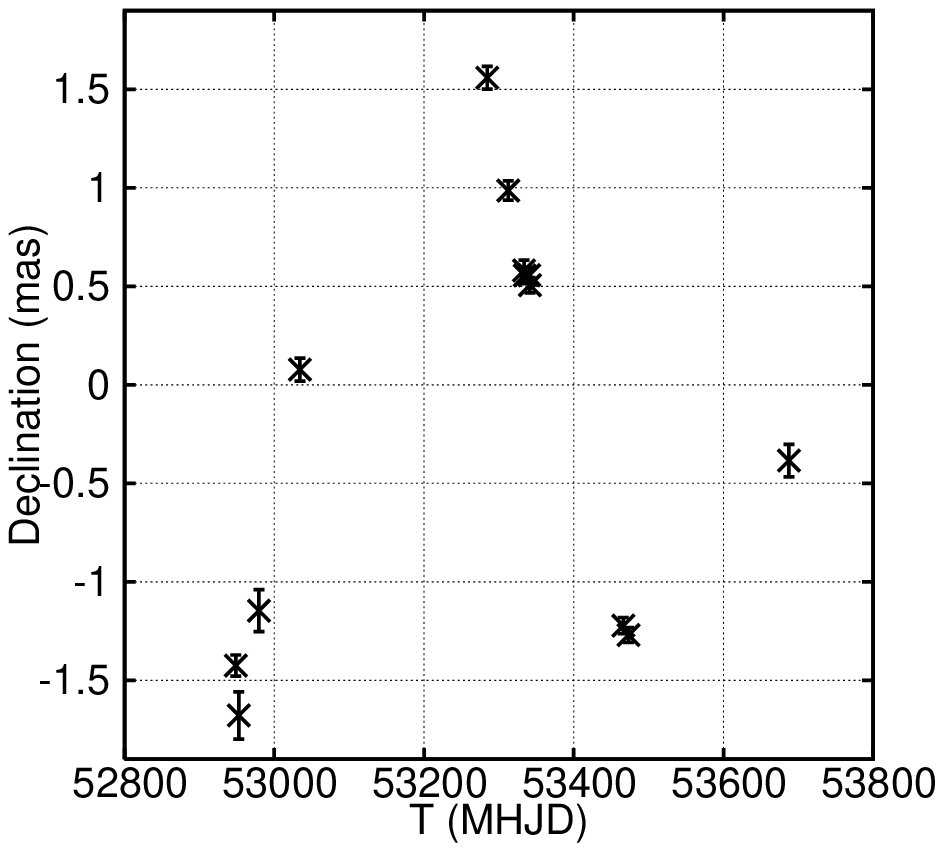}
\caption[HR 2896 Binary-Only Fit Residuals]
{ \label{fig::hr2896_binary_resid}
Residuals to a single Keplerian fit to the HR 2896 astrometry measurements, 
in the 
declination axis.  (Left) Both PHASES (X symbols)and non-PHASES ($+$ symbols) 
measurements are shown.  (Right) For clarity, only PHASES measurements are 
shown.  The presence of a long period tertiary companion is obvious in the 
PHASES measurements.  A refined search confirmed the signal is Keplerian with 
a period of $\sim 700$ days.
}
\end{figure*}

While the orbital periods of the second Keplerian perturbations were relatively 
obvious either from previous spectroscopic work (63 Gem A) or visual inspection 
of residuals to a single Keplerian model (HR 2896; see Figure 
\ref{fig::hr2896_binary_resid}), 
blind searches were conducted to ensure that aliased orbital periods were 
not being misidentified.  An algorithm, based on that of \cite{cumming1999} and 
\cite{cumming2008} was modified for use on astrometric data for binary systems 
(described in Paper III) and used to conduct blind searches for 
tertiary companions in both systems.

The overall procedure is to create a periodogram of an 
F statistic comparing the goodness-of-fit $\chi^2$ between a single Keplerian 
model and that for a double Keplerian model for a number of possible orbital 
periods for the second orbit.  The orbital periods selected were chosen to be 
more than Nyquist sampled, to ensure complete coverage, as $P = 2 f T / k$
where $T$ is the span of PHASES observations, $f$ is an oversampling factor, 
and $k$ is a positive integer.  Two searches were 
conducted for 63 Gem:  first, using only the PHASES measurements, and 
second using both the PHASES and non-PHASES astrometry, to better constrain the 
wide binary motion during the search.  Only the first of these searches was 
conducted for HR 2896, since the signal was very large.  
For both searches of 63 Gem A, $f=2$ was 
chosen for computational efficiency, because the largest value of $k$ was much 
larger than typical, corresponding to that for which the minimum period 
sampled was 1.1 days, to ensure inclusion of the expected $\lesssim 2$ day 
companion orbital period, and only positive integer values of $k$ were 
evaluated.  For the search of HR 2896, $f=3$ was chosen, and the largest 
value of $k$ corresponded to that for which the minimum orbital period examined 
was 6 days.  In addition to the positive integer values of $k$, the period 
corresponding to $k=1/2$ was evaluated to search for companions with orbits 
slightly longer than the PHASES span.

The orbital period for which the F statistic periodogram has its maximum value 
is the most likely orbital period of a companion object.  To ensure the peak is 
a real object rather than a statistical fluctuation, 1000 synthetic data sets 
with identical cadence and measurement uncertainties as the actual data were 
created and evaluated in the same manner.  The fraction of these having a 
maximum F statistic larger than that of the actual data provided an estimate of 
the false alarm probability (FAP) that the signal is not caused by an actual 
companion.

\begin{figure*}[!ht]
\plottwo{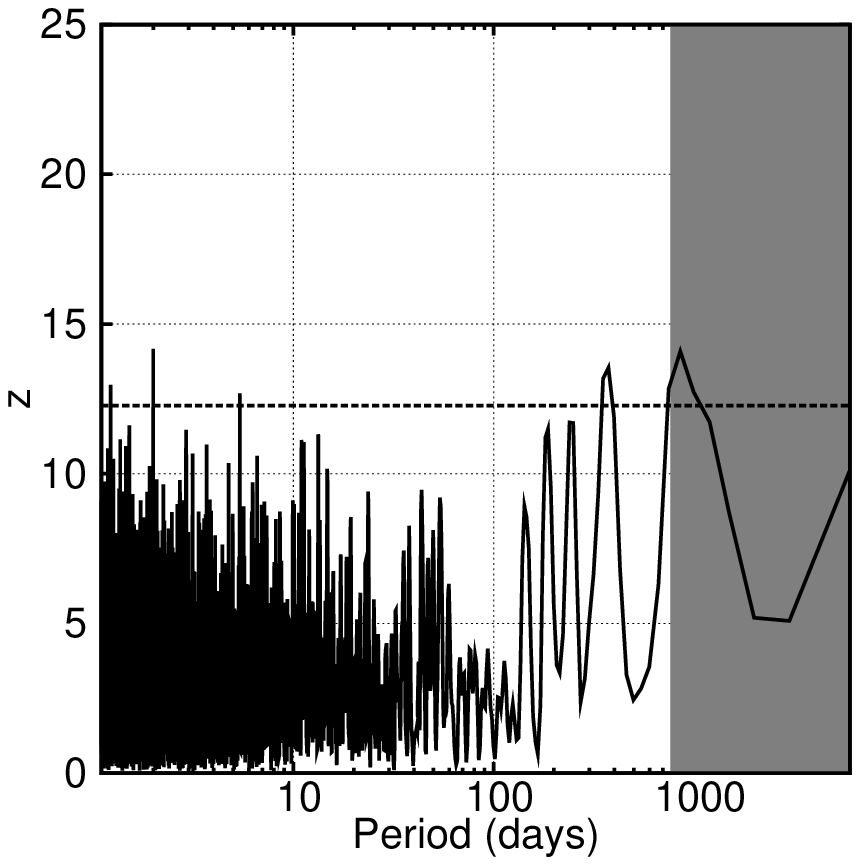}{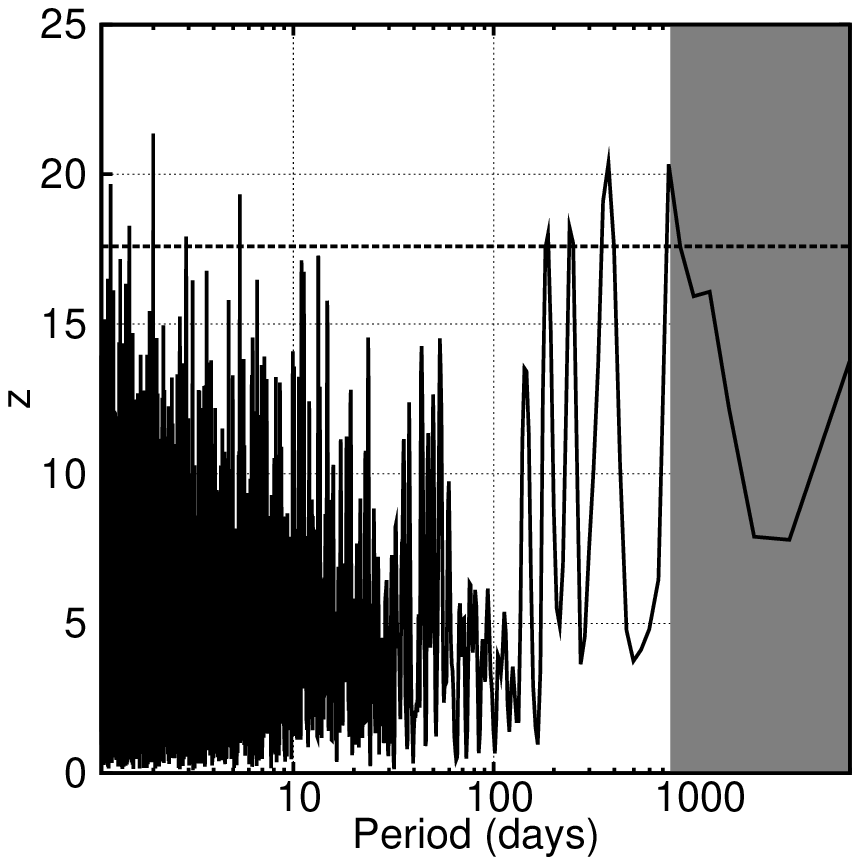}
\caption[Periodograms for Tertiary Companions to 63 Gem A and HR 2896]
{ \label{fig::periodograms_58728}
Periodograms of the F statistic (discussed in the text and written here as 
$z$) for 63 Gem A.  The larger the value of z, the more likely the
data are to represent the presence of a third component at the given
period.  Orbital periods longer than 760 days are unphysical, as this is the 
orbital period of the wider system itself; these regions are shaded in the 
plots.  The left 
figure is for analysis only using the PHASES data, while the right is for 
combined analysis of PHASES and non-PHASES astrometric measurements.
For 63 Gem A the 99\% confidence level is at $z=12.3$ for the PHASES-only 
analysis, and $z=17.6$ for the combined analysis, as indicated by horizontal 
lines.
}
\end{figure*}

Because the 63 Gem Aa1-Aa2 subsystem has 
been well-established by RV observations, it 
was not necessary that the astrometric data meet ``detection'' criteria 
to simply place constraints on an orbit.  
However, it is interesting to note a wobble was detected in the PHASES 
data at a period near that of the RV signal, though not exactly overlapping.  
The peak in the 63 Gem A periodogram was at a period of 1.997 days in both 
the PHASES-only and combined astrometry searches; for the former, the 
peak value was $z=14.17$ with an FAP of 0.3\%, whereas for the latter this 
improved to $z=21.36$ with an FAP of 0.1\%.  However, the much higher 
signal-to-noise ratio  
RV signal was found to have an orbital period closer to 1.93 days; the 
PHASES measurements are consistent with this signal as well, though the 
search algorithm did not exactly reproduce this value.  This is likely 
due to the low signal level of the astrometric perturbation, and aliasing 
issues related to observing cadences.  The periodograms of the 63 Gem data 
are presented in Figure \ref{fig::periodograms_58728}.

\begin{figure}[!ht]
\plotone{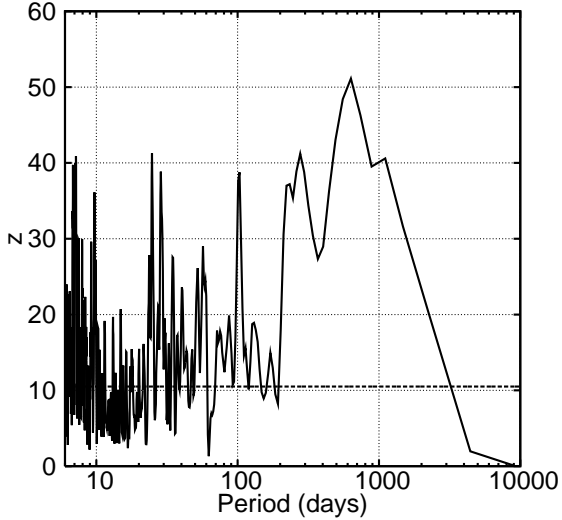}
\caption[Periodogram for Tertiary Companions to HR 2896]
{ \label{fig::periodograms_60318}
Periodogram of the F statistic (discussed in the text and written here as 
$z$) for HR 2896.  The larger the value of z, the more likely the
data are to represent the presence of a third component at the given
period.  For HR 2896 the 99\% confidence level is at $z=10.5$, indicated by 
a horizontal line.
}
\end{figure}

For HR 2896, the 99\% confidence level was at $z=10.5$.  With a maximum 
$z=51.1$ at $P = 633$ days, none of the 1000 trials had $z$ values exceeding 
that of the data; the FAP is estimated to be less than 0.1\%.  The 633 day 
orbital period was refined when the full double Keplerian model was 
fit to all the astrometric data.  The periodogram of the HR 2896 data 
is presented in Figure \ref{fig::periodograms_60318}.

\section{Orbital Models}

\subsection{Orbit Fitting}

Measurements were fit to models consisting of two Keplerian orbits 
superimposed on each other.  One represented the motion of the wide system, 
and the other that of the subsystem.  The best-fit orbit parameters for the 
63 Gem A and HR 2896 triple systems are listed in Table \ref{tab::OrbitModels}.

\begin{figure}[!ht]
\plotone{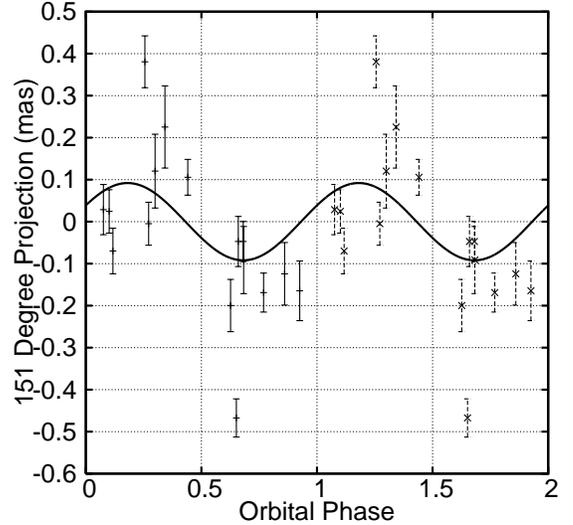}
\caption[Center-Of-Light Orbit of 63 Gem Aa1-Aa2]
{ \label{fig::63GemOrbit}
The center of light orbit of the $\sim 1.9$ day 63 Gem Aa1-Aa2 subsystem, 
along an axis at angle 151 degrees, measured from 
increasing differential right ascension through increasing differential 
declination; it was along this axis that the PHASES measurements were typically 
most sensitive.  For clarity, only measurements with uncertainties along this 
axis of $< 100\, {\rm \mu as}$ are shown, and the measurements 
have been phase-wrapped about the 1.9 day orbital period and double plotted 
to cover two cycles.
}
\end{figure}

\begin{figure*}[!ht]
\plottwo{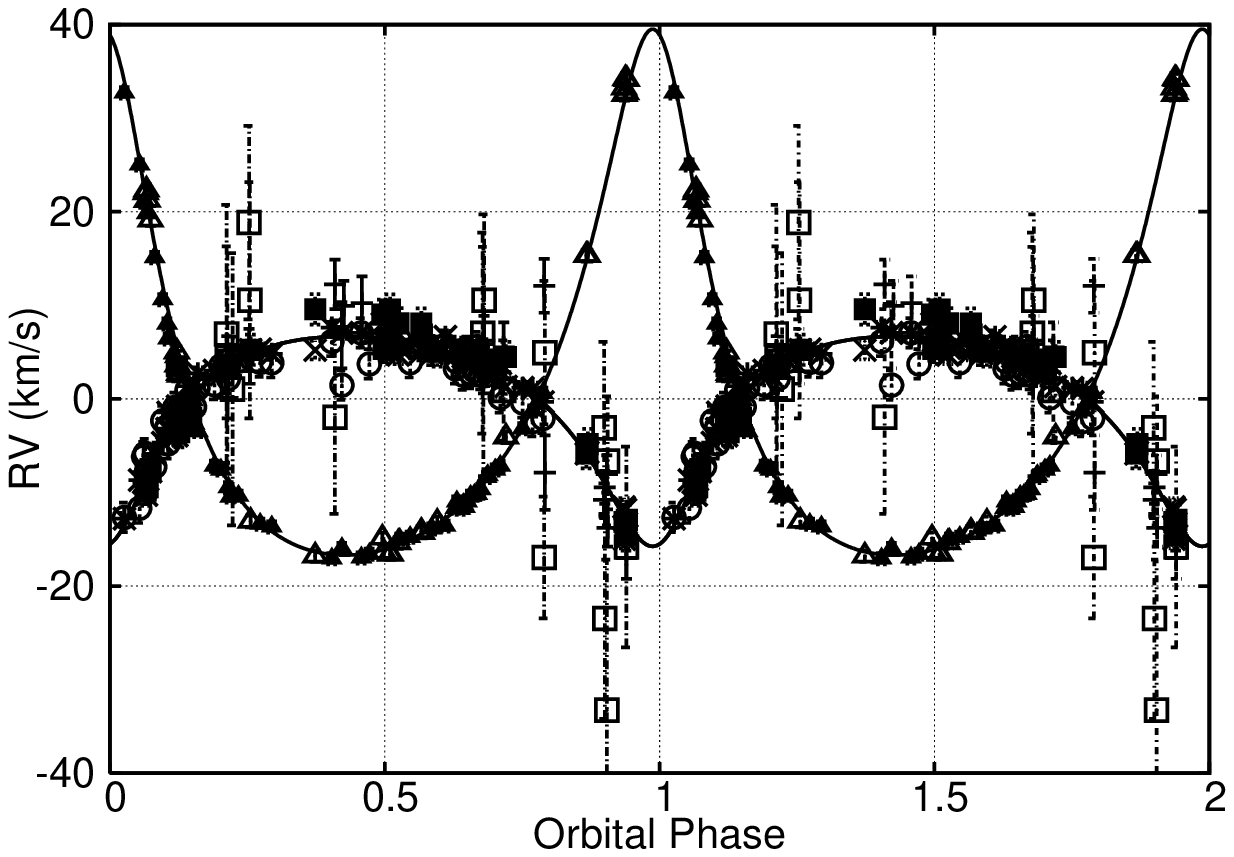}{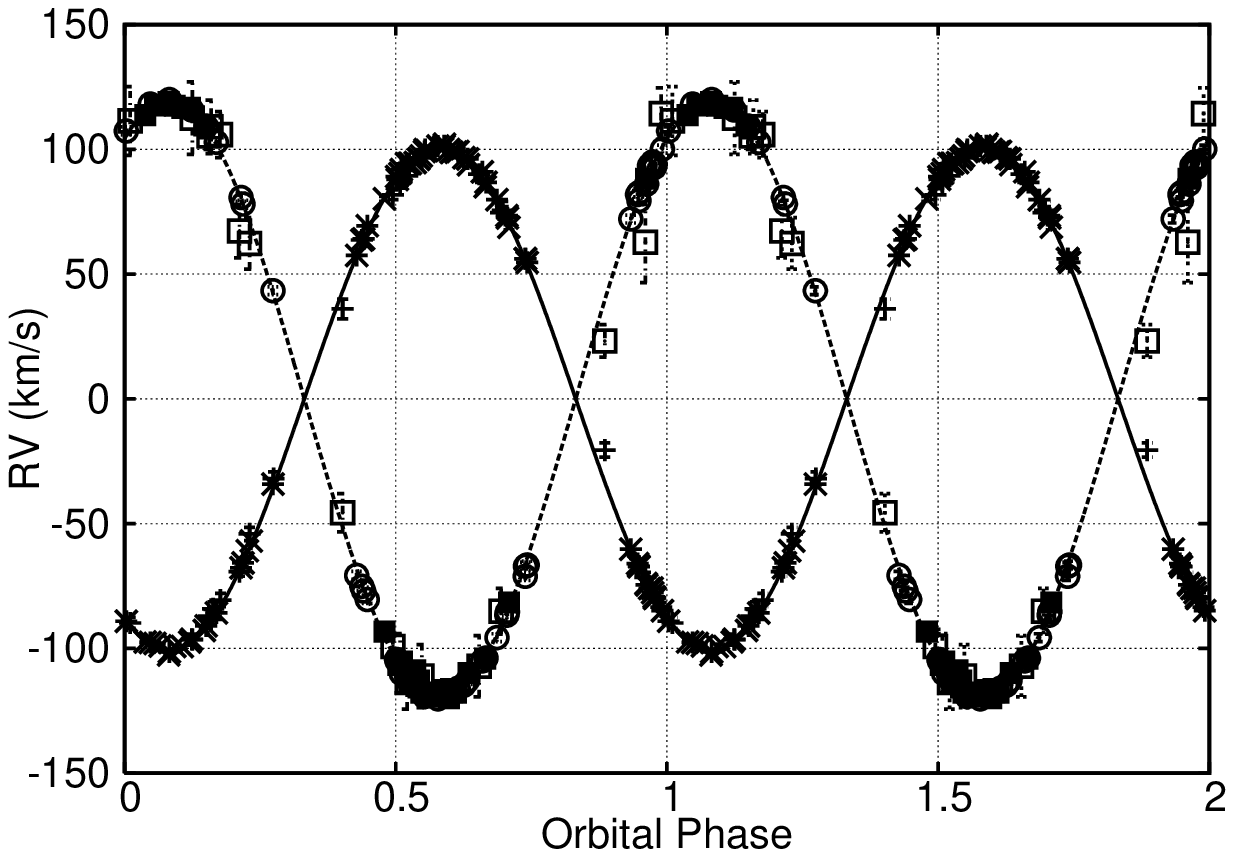}
\caption[Radial Velocity Orbit of 63 Gem Aa-Ab and Aa1-Aa2]
{ \label{fig::63GemRV}
(Left) 
Radial velocity orbit of 63 Gem Aa-Ab, the outer pair.  For clarity, the 
system velocity offset and Aa1-Aa2 signals have been removed, and the data were 
phase-wrapped about the 760-day orbital period.  Two cycles of the orbit 
are shown for continuity (the measurements are double-plotted).
(Right) Radial velocity orbit of the 63 Gem Aa1-Aa2 subsystem.  Again the 
system velocity offset and Aa-Ab signal have been removed, and the data were 
phase-wrapped about the 1.9-day orbital period and double-plotted.  
Measurements by \cite{abt1976} are marked by $+$ and unfilled square symbols 
for components Aa1 and Aa2, respectively.  Measurements from Kitt Peak are 
marked by $\times$, filled square, and open triangle symbols for components 
Aa1, Aa2, and Ab, respectively.  Measurements from the AST are marked by 
asterisks, open circles, and filled triangle symbols for components 
Aa1, Aa2, and Ab, respectively.
}
\end{figure*}

For 63 Gem A, with velocities of all three 
components, the total mass of Aa1-Aa2 
can be determined by its motion in the wide Aa-Ab system and the part of its 
velocity from that.  Coupled with the Aa1-Aa2 orbital period, this mass 
constrained the semimajor axis of the subsystem.  When the Aa1 and Aa2 velocity 
signals from the subsystem motion were analyzed, this in turn constrained the 
inclination of the Aa1-Aa2 subsystem.  Thus, just from a visual orbit of the 
outer system, and radial velocities of all components, all parameters of the 
inner system are constrained except the longitude of the ascending node 
$\Omega_{\rm Aa1-Aa2}$.  The high precision PHASES astrometric measurements were 
necessary to constrain this one last parameter, crucial for understanding the 
overall system geometry.  The Aa1-Aa2 center of light motion is plotted in 
Figure \ref{fig::63GemOrbit} and the Aa1-Aa2 and Aa-Ab radial velocity orbits 
are plotted in Figure \ref{fig::63GemRV}.

\begin{figure*}[!ht]
\plottwo{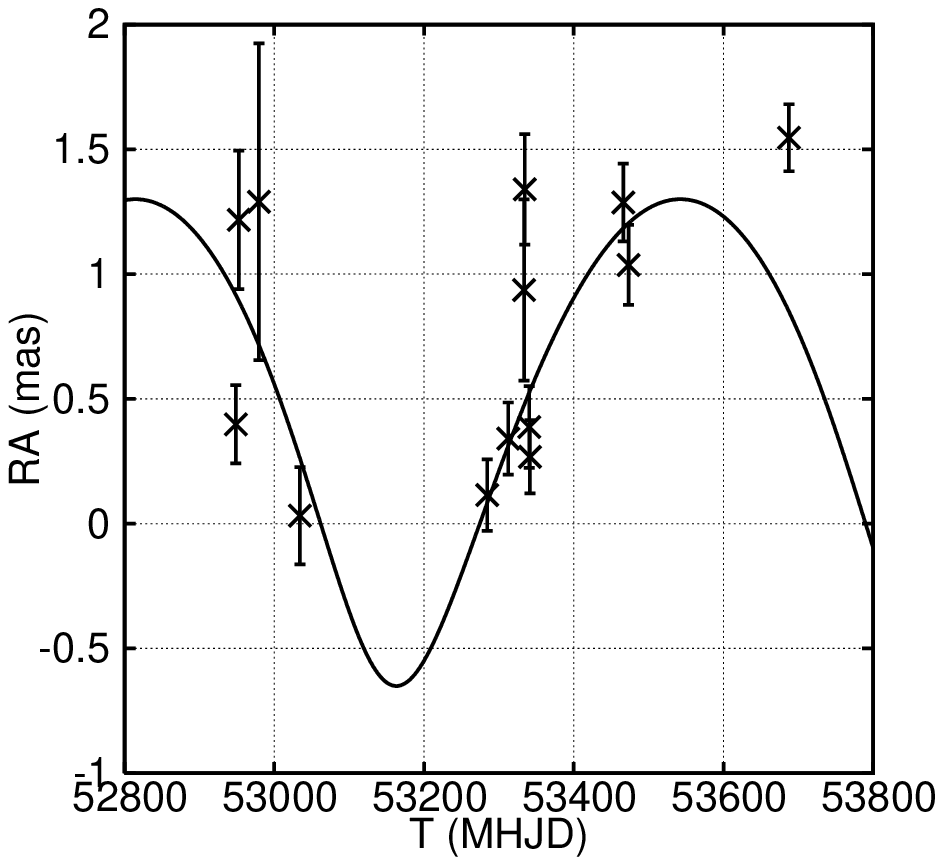}{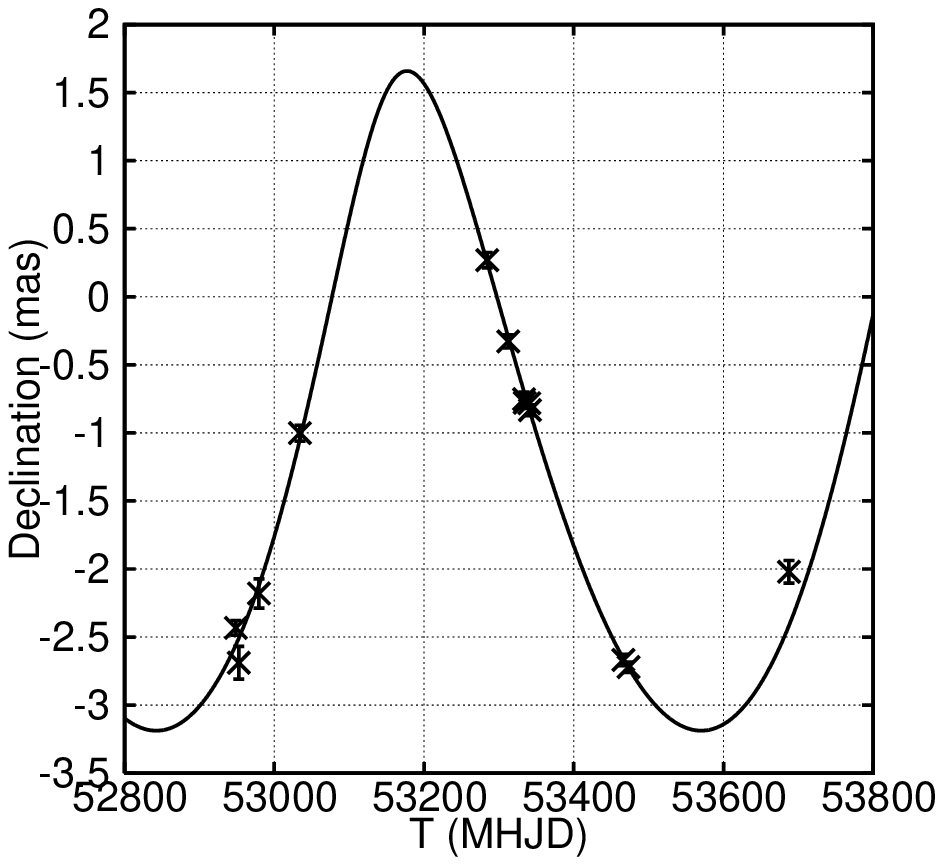}
\caption[Center-Of-Light Orbit of the HR 2896 Subsystem]
{ \label{fig::hr2896Orbit}
Motion of the center-of-light of the 730 day subsystem in HR 2896.  (Left) the 
motion in the Right Ascension axis, which 
was typically lower precision for the 
single baseline interferometric measurements (which were most often taken with 
the nearly North-South baseline at PTI).  (Right) The motion in the Declination 
axis.
}
\end{figure*}

For HR 2896, only astrometric measurements are available.
From astrometry alone, it is impossible to tell which component contains the 
astrometric subsystem.  In evaluating the orbital model parameters, it is 
assumed the secondary is the subsystem, though this is an arbitrary selection.  
The center-of-light motion of the HR 2896 subsystem is plotted in Figure 
\ref{fig::hr2896Orbit}.

Without a second Keplerian, HR 2896 fit a single Keplerian model with 
$\chi^2 = 8037.7$ and 381 degrees of freedom, for a reduced $\chi_r^2 = 21.1$.  
The double Keplerian model represented the data much better:  $\chi^2 = 432.6$ 
with 374 degrees of freedom and $\chi_r^2 = 1.16$.  

The orbital period of the subsystem in HR 2896 is almost exactly two years.  
While this can be a source of some concern about systematic effects, the large 
signal, its presence in two dimensions, and the fact that it does not appear 
in any of the other PHASES binaries lessens this concern.  Coincidentally, 
the period of the outer pair of 63 Gem A (760 days) is nearly equal to that of 
the inner pair in HR 2896 (730 days); this demonstrates the great variety of 
configurations of hierarchical triple star systems.

\begin{deluxetable*}{lllll}
\tablecolumns{5}
\tablewidth{0pc} 
\tablecaption{Orbit models for 63 Gem A and HR 2896\label{tab::OrbitModels}}
\tablehead{
\colhead{} & \multicolumn{2}{c}{63 Gem A} & \multicolumn{2}{c}{HR 2896} \\
\colhead{Parameter} & \colhead{Value} & \colhead{Uncertainty} & \colhead{Value} & \colhead{Uncertainty} 
}
\startdata 
$P_{Wide}$ (days)                             
& 760.083   & $\pm 0.081$     & 77931   & $\pm 1097$  \\
$T_{Wide}$ (MHJD)                             
& 54038.6   & $\pm 1.3 $      & 44117   & $\pm 93$    \\
$e_{Wide}$                                    
& 0.4150    & $\pm 0.0014$    & 0.6756  & $\pm 0.0051$ \\
$a_{Wide}$ (arcsec)                     
& \nodata   & \nodata         & 0.5554  & $\pm 0.0075$   \\
$i_{Wide}$ (degrees)                          
& 92.31     & $\pm 0.12  $    & 91.788  & $\pm 0.042$ \\
$\omega_{Wide}$ (degrees)                     
& 192.15    & $\pm 0.59 $     & 137.5   & $\pm 1.1$   \\
$\Omega_{Wide}$ (degrees)                     
& 346.930   & $\pm 0.060$     & 329.89  & $\pm 0.14$  \\
 $P_{\rm subsystem}$ (days)                           
& 1.93267835 & $\pm 4.2 \times 10^{-6}$   
& 727.9     & $\pm 8.6              $ \\
$T_{\rm subsystem}$ (MHJD)                           
& 54051.14  & $\pm 0.46$      & 53876    & $\pm 19$   \\
$e_{\rm subsystem}$                                  
& 0.0012    & $\pm 0.0019$    & 0.339    & $\pm 0.074$ \\
$a_{{\rm COL}, \, Ba-Bb}$ (arcsec)                     
& \nodata   & \nodata         & 0.00263  & $\pm 0.00017$   \\
$i_{\rm subsystem}$ (degrees)                        
& 69.9      & $\pm 1.2$       & 86.2     & $\pm 1.6$  \\
$\omega_{\rm subsystem}$ (degrees)                   
& 151       & $\pm 86 $       & 338.6    & $\pm 6.6$  \\
$\Omega_{\rm subsystem, 1}$ (degrees)                   
& 145       & $\pm 16$        & 158.5    & $\pm 2.8$  \\
$\Omega_{\rm subsystem, 2}$ (degrees)                   
& 325       & $\pm 16$        & 338.5    & $\pm 2.8$  \\
$V_{0, {\rm Abt \, \& \, Levy}}$ (${\rm km~s^{-1}}$)               
& 24.57     & $\pm 0.81$      & \nodata & \nodata     \\
$V_{0, {\rm Kitt\, Peak}}$ (${\rm km~s^{-1}}$)               
& 22.70     & $\pm 0.16$      & \nodata & \nodata     \\
$V_{0, {\rm AST}}$ (${\rm km~s^{-1}}$)               
& 22.56     & $\pm 0.11$      & \nodata & \nodata     \\
$M_{Aa1+Aa2}$ ($\Msun$)                        
& 2.583     & $\pm 0.059$     & \nodata & \nodata     \\
$M_{Aa2}/M_{Aa1}$                              
& 0.8422    & $\pm 0.0029$    & \nodata & \nodata     \\
$L_{Aa2}/L_{Aa1}$ (Option 1)                   
& 0.469     & $\pm 0.099$     & \nodata & \nodata     \\
$L_{Aa2}/L_{Aa1}$ (Option 2)                             
& 1.47      & $\pm 0.28$      & \nodata & \nodata     \\
$M_{Ab}$ ($\Msun$)                              
& 1.030     & $\pm 0.038$     & \nodata & \nodata     \\
$d$ (parsecs)                                
& 33.09     & $\pm 0.28$      & \nodata & \nodata     \\
$\chi^2$ and DOF 
& 926.1     & 332             & 432.6   & 374         \\
\tableline
$\Phi_1$ (degrees)               
& 152       & $\pm 12 $       & 10.2    & $\pm 2.4$   \\
$\Phi_2$ (degrees)               
& 31        & $\pm 11 $       & 171.2   & $\pm 2.8$   \\
$M_{Aa1}$ ($\Msun$)                           
& 1.402     & $\pm 0.032$     & \nodata & \nodata     \\
$M_{Aa2}$ ($\Msun$)                           
& 1.181     & $\pm 0.027$     & \nodata & \nodata     \\
$M_{Ba}$ ($\Msun$)                           
& \nodata   & \nodata         & 1.3     & \nodata     \\
$M_{Bb}$ ($\Msun$)                           
& \nodata   & \nodata         & 0.2     & \nodata     \\
$a_{Wide}$ (arcsec)                           
& 0.07558   & $\pm 0.00026$   & \nodata & \nodata     \\
$a_{\rm Aa1-Aa2}$ (arcsec)                             
& 0.0005973 & $\pm 0.0000089$ & \nodata & \nodata     \\
$a_{Wide}$ (AU)                           
& 2.501     & $\pm 0.016$     & 51.5    & $\pm 5.6$   \\
$a_{\rm subsystem}$ (AU)                             
& 0.04166   & $\pm 0.00032$   & 1.81    & $\pm 0.20$  \\
$\pi$ (mas)                                  
& 30.22     & $\pm 0.26$      & \nodata & \nodata \
\enddata
\tablecomments{Best fit orbit parameters for the triple systems 
63 Gem A and HR 2896.  Quantities below the line were derived, 
and their uncertainties estimated using first order error 
propagation.
}
\end{deluxetable*}

\subsection{Derived Quantities}

For HR 2896, the {\em Hipparcos} parallax of $10.78 \pm 1.16$ mas was used to 
set an overall physical scale size for the system; this uncertainty was added 
in quadrature in the standard manner for first-order error propagation when 
calculating other derived quantities' uncertainties.  From the binary (wide 
system) orbital period, semimajor axis, and the parallax, the total system mass 
is found to be $3.00 \pm 0.98 \, \Msun$.  Without a mass ratio, it was assumed 
this mass is divided evenly between components (that is, the 730 day subsystem 
has total mass $1.50 \, \Msun$, equal to that of the other 
component).  While this is very inexact, it nonetheless allows an estimate of 
the mass and nature of the unseen perturber.

Assuming a total mass of 
$1.50 \, \Msun$, the 730 day orbital period, center of light semimajor 
axis (taken to be the true deflection of the more luminous component of 
the subsystem, an 
approximation justified when the perturber's mass was determined), 
and the parallax, a mass of $0.202 \, \Msun$ was derived.  
The approximation that the luminosity of 
the perturber can be ignored was justified since such a low mass star would be 
far fainter than a K giant.  Furthermore, even if the subsystem's total mass 
were assigned to be $3.98 \, \Msun$ (the 1-$\sigma$ upper bound of the total 3 
body mass), the mass of the perturber would be 
only $0.39 \, \Msun$, still in the 
range of M dwarfs and still far less luminous than a K giant.

HR 2896 is likely to be in a stable configuration.  Based on a large sample of 
gravitational simulations of test masses in binary systems, \cite{holman1999} 
formulated an empirical criteria for whether a tertiary companion can be 
stable long term.  Their formula indicates objects with periods as long as 
2100 days would be stable in HR 2896.

From the total subsystem mass approximation of $1.50 \pm 0.49 \, \Msun$ and 
730 day orbital period, the full semimajor axis of the subsystem is 
$19.5 \pm 3.0$ mas.  This is half the resolution of the {\em Hubble Space 
Telescope} and most infrared adaptive optics systems, but is easily in range 
of long baseline interferometry systems if high contrast ratios are possible.  
The high-precision closure phases being obtained by the MIRC beam combiner at 
CHARA \citep{MonnierMirc, CHARAEXISTS} might enable imaging of the third 
component.  Since it is likely the tertiary is a very red star, infrared 
imaging is likely to be advantageous for reducing the contrast requirements.  
From the parallax, it is found that the wider binary has a semimajor axis of 
$51.5 \pm 5.6$ AU and the subsystem has a semimajor axis of $1.81 \pm 0.20$ 
AU.  Thus, the high eccentricity of the wide pair ($0.6756 \pm 0.0051$) brings 
the outer star to within a factor of 10 of the size of the subsystem itself.

It is impossible to discern between two possible values of the mutual 
inclination of the two orbits from astrometry alone, because one must identify 
which node is ascending.  From astrometry, two orbital solutions are possible, 
separated by changing both $\Omega$ and $\omega$ by $180^\circ$ in either (or 
both) orbits.  The mutual inclination of the orbits is given by
\begin{equation}\label{KapPegMutualInclination}
\cos \Phi = \cos i_1 \cos i_2  + \sin i_1 \sin i_2 \cos\left(\Omega_1 - \Omega_2\right)
\end{equation}
\noindent where $i_1$ and $i_2$ are the orbital inclinations and $\Omega_1$ 
and $\Omega_2$ are the longitudes of the ascending nodes.  For HR 2896, the two 
possible values are $10.2 \pm 2.4$ degrees or $171.2 \pm 2.8$ degrees.  These 
values are very close to being either coplanar or anticoplanar; in either case, 
this is relatively rare for a triple star system.

Even when a center-of-light astrometric orbit and radial velocities are 
available for the narrow 
pair, there can be a degeneracy between which node is ascending and the 
luminosity ratio.  Having found one possible luminosity ratio 
$L_{\rm Aa2}/L_{\rm Aa1} = L_1$, it can be shown that the other possible 
solution, corresponding to varying the ascending node by 180 degrees, is given 
by
\begin{equation}\label{eqLum}
L_2 = \frac{2R+RL_1-L_1}{1+2L_1-R}
\end{equation}
\noindent where $R$ is the mass ratio $M_{\rm Aa2}/M_{\rm Aa1}$.

For 63 Gem A it is possible to lift the degeneracy, because the solution 
for which $L_{\rm Aa2}/L_{\rm Aa1} = 1.47 > 1$ is inconsistent with the spectral 
line strengths.  Thus, the 
mutual inclination was determined without ambiguity to be $152 \pm 12$ 
degrees.  This is near the range $39.2-140.8$ degrees where Kozai 
oscillations between inner pair orbital inclination and eccentricity are 
driven \citep{Kozai1962}.  
Furthermore, the large mutual inclination again shows that star systems 
tend not to 
be coplanar, unlike the planets of the solar system.  Should other planetary 
systems also be preferentially coplanar, this suggests an important 
difference in the modes and timescales of planet versus binary formation.

\section{Discussion}

\subsection{63 Gem A}

With three luminous components, all of which can be studied spectroscopically, 
63 Gem A is a valuable system for triple star studies.  The short period 
subsystem's very small separation (0.6 mas) would push the resolution limits 
of even long baseline interferometry.  Because the contrast is low, a system 
operating at visible wavelengths (where spatial resolution is enhanced) and 
capable of precision closure phases (for imaging) could be used to continue 
studying this compelling system.  The recently NSF funded Visible Imaging 
System for Interferometric Observations at NPOI (VISION) beam combiner will 
contribute to this effort.

\vspace{0.5in}
\subsection{HR 2896}

The low mass companion to HR 2896 is likely a M dwarf in either a nearly 
coplanar or anti-coplanar orbit to the wide binary itself.  While a 
$0.2 \, \Msun$ white dwarf is also a possibility, this seems unlikely; such a 
low mass white dwarf could only be produced by a low mass star that would be 
unlikely to have evolved to that point by now, and almost certainly not before 
the more massive stars in the system would have evolved.  With an anticipated 
separation of 19.5 mas, the M dwarf could easily be resolved by long baseline 
interferometry, though the high contrast will be challenging.  An instrument 
operating at infrared wavelengths, where the contrast is lower, and capable of 
precision closure phases, such as MIRC at the CHARA Array, could be used for 
further study of this system.

{\it Facilities:} \facility{PO:PTI, TSU:AST, KPNO:CFT}

\acknowledgements 
PHASES benefits from the efforts of the PTI collaboration members who have 
each contributed to the development of an extremely reliable observational 
instrument.  Without this outstanding engineering effort to produce a solid 
foundation, advanced phase-referencing techniques would not have been 
possible.  We thank PTI's night assistant Kevin Rykoski for his efforts to 
maintain PTI in excellent condition and operating PTI in phase-referencing 
mode every week.  Thanks are also extended to Ken Johnston and the 
U.~S.~Naval Observatory for their continued support of the USNO Double Star 
Program.  Part of the work described in this paper was performed at 
the Jet Propulsion Laboratory under contract with the National Aeronautics 
and Space Administration.  Interferometer data was obtained at the Palomar
Observatory with the NASA Palomar Testbed Interferometer, supported
by NASA contracts to the Jet Propulsion Laboratory.  This publication makes 
use of data products from the Two Micron All Sky Survey, which is a joint 
project of the University of Massachusetts and the Infrared Processing and 
Analysis Center/California Institute of Technology, funded by the National 
Aeronautics and Space Administration and the National Science Foundation.  
This research has made use of the Simbad database, operated at CDS, 
Strasbourg, France.  This research has made use of SAOImage DS9, developed 
by the Smithsonian Astrophysical Observatory.  
MWM acknowledges support from the Townes Fellowship 
Program, Tennessee State University, and the state of Tennessee through its 
Centers of Excellence program.  Some of the software used for analysis was 
developed as part of the SIM Double Blind Test with support from NASA 
contract NAS7-03001 (JPL 1336910).  
PHASES is funded in part by the California Institute of Technology 
Astronomy Department, and by the National Aeronautics and Space Administration 
under Grant No.~NNG05GJ58G issued through the Terrestrial Planet Finder 
Foundation Science Program.  This work was supported in part by the National 
Science Foundation through grants AST 0300096, AST 0507590, and AST 0505366.
MK is supported by the Foundation for Polish Science through a FOCUS 
grant and fellowship, by the Polish Ministry of Science and Higher 
Education through grant N203 3020 35.

\bibliography{main}
\bibliographystyle{apj}

\clearpage
\LongTables 
\begin{deluxetable*}{llrrrrll}
\tablecolumns{8}
\tablewidth{0pc} 
\tablecaption{Non-PHASES Astrometric Measurements \label{tab::speckleData}}
\tablehead{ 
\colhead{HD Number} & \colhead{Date} & \colhead{$\rho$} & \colhead{$\theta$} 
& \colhead{$\sigma_{\rho}$} & \colhead{$\sigma_{\theta}$} & \colhead{Weight} 
& \colhead{Outlier}\\
\colhead{} & \colhead{(Year)} & \colhead{(arcsec)} & \colhead{(degrees)} 
& \colhead{(arcsec)} & \colhead{(degrees)} & \colhead{} & \colhead{}}
\startdata
58728 & 1980.1588 & 0.044 & 348.00 & 0.013 & 4.29 & 1.2 & 0 \\
58728 & 1983.0476 & 0.114 & 346.60 & 0.005 & 1.64 & 8.2 & 0 \\
58728 & 1983.9554 & 0.035 & 163.00 & 0.015 & 4.95 & 0.9 & 0 \\
58728 & 1983.9583 & 0.036 & 171.50 & 0.015 & 4.70 & 1.0 & 0 \\
58728 & 1984.8436 & 0.109 & 347.30 & 0.005 & 1.64 & 8.2 & 0 \\
58728 & 1985.0000 & 0.104 & 347.20 & 0.005 & 1.75 & 7.2 & 0 \\
58728 & 1985.1801 & 0.089 & 345.00 & 0.006 & 1.99 & 5.6 & 0 \\
58728 & 1985.2050 & 0.071 & 354.50 & 0.046 & 14.86 & 0.1 & 0 \\
58728 & 1986.8867 & 0.103 & 346.80 & 0.006 & 1.76 & 7.1 & 0 \\
58728 & 1986.8893 & 0.104 & 347.10 & 0.005 & 1.75 & 7.2 & 0 \\
58728 & 1987.2689 & 0.094 & 346.30 & 0.006 & 1.95 & 5.8 & 0 \\
58728 & 1988.1648 & 0.047 & 181.20 & 0.014 & 4.48 & 1.1 & 0 \\
58728 & 1988.2520 & 0.089 & 199.20 & 0.006 & 2.02 & 5.4 & 1 \\
58728 & 1988.9098 & 0.106 & 350.00 & 0.018 & 5.62 & 0.7 & 0 \\
58728 & 1989.1556 & 0.107 & 346.70 & 0.012 & 3.84 & 1.5 & 0 \\
58728 & 1989.2295 & 0.102 & 346.70 & 0.006 & 1.76 & 7.1 & 0 \\
58728 & 1993.2025 & 0.105 & 341.40 & 0.008 & 2.44 & 3.7 & 0 \\
58728 & 1993.8396 & 0.061 & 350.60 & 0.008 & 2.71 & 3.0 & 0 \\
58728 & 1994.8707 & 0.071 & 350.30 & 0.009 & 2.86 & 2.7 & 0 \\
58728 & 1995.1490 & 0.094 & 347.00 & 0.010 & 3.10 & 2.3 & 0 \\
58728 & 1995.3158 & 0.093 & 342.20 & 0.010 & 3.10 & 2.3 & 0 \\
58728 & 1996.8636 & 0.087 & 340.70 & 0.011 & 3.41 & 1.9 & 0 \\
58728 & 1997.0740 & 0.078 & 1.00 & 0.023 & 7.43 & 0.4 & 0 \\
58728 & 1997.1259 & 0.094 & 348.10 & 0.008 & 2.59 & 3.3 & 0 \\
58728 & 1997.8274 & 0.074 & 350.10 & 0.012 & 3.72 & 1.6 & 0 \\
58728 & 1997.8274 & 0.072 & 352.30 & 0.012 & 3.84 & 1.5 & 0 \\
58728 & 1999.1602 & 0.111 & 346.30 & 0.016 & 5.26 & 0.8 & 0 \\
58728 & 2007.8230 & 0.114 & 347.80 & 0.008 & 2.48 & 3.6 & 0 \\
58728 & 2007.8260 & 0.105 & 348.90 & 0.008 & 2.63 & 3.2 & 0 \\
60318 & 1842.9500 & 0.500 & 331.20 & 0.101 & 3.56 & 0.7 & 0 \\
60318 & 1843.2400 & 0.420 & 340.90 & 0.267 & 9.42 & 0.1 & 0 \\
60318 & 1845.2700 & 0.470 & 332.60 & 0.109 & 3.85 & 0.6 & 0 \\
60318 & 1846.3000 & 0.450 & 333.50 & 0.267 & 9.42 & 0.1 & 0 \\
60318 & 1848.2900 & 0.440 & 329.00 & 0.101 & 3.56 & 0.7 & 0 \\
60318 & 1851.3300 & 0.380 & 334.90 & 0.267 & 9.42 & 0.1 & 0 \\
60318 & 1852.2500 & 0.460 & 331.40 & 0.134 & 4.71 & 0.4 & 0 \\
60318 & 1854.2800 & 0.500 & 329.40 & 0.267 & 9.42 & 0.1 & 0 \\
60318 & 1861.1899 & 0.620 & 328.80 & 0.120 & 4.21 & 0.5 & 0 \\
60318 & 1869.5000 & 0.850 & 331.30 & 0.071 & 2.52 & 1.4 & 0 \\
60318 & 1879.2500 & 0.720 & 334.50 & 0.101 & 3.56 & 0.7 & 0 \\
60318 & 1879.7560 & 0.610 & 331.90 & 0.053 & 1.88 & 2.5 & 0 \\
60318 & 1880.2000 & 0.900 & 331.70 & 0.154 & 5.44 & 0.3 & 0 \\
60318 & 1883.4800 & 0.880 & 332.20 & 0.089 & 3.14 & 0.9 & 0 \\
60318 & 1884.0400 & 0.750 & 332.80 & 0.040 & 1.42 & 4.4 & 0 \\
60318 & 1885.0400 & 0.780 & 330.60 & 0.109 & 3.85 & 0.6 & 0 \\
60318 & 1888.2380 & 0.710 & 334.10 & 0.051 & 1.78 & 2.8 & 0 \\
60318 & 1889.9700 & 0.720 & 330.60 & 0.048 & 1.69 & 3.1 & 0 \\
60318 & 1891.1479 & 0.690 & 329.70 & 0.085 & 2.98 & 1.0 & 0 \\
60318 & 1893.1400 & 0.820 & 330.40 & 0.120 & 4.21 & 0.5 & 0 \\
60318 & 1893.2500 & 0.500 & 329.90 & 0.154 & 5.44 & 0.3 & 0 \\
60318 & 1894.2100 & 0.810 & 334.90 & 0.134 & 4.71 & 0.4 & 0 \\
60318 & 1895.2500 & 0.730 & 332.40 & 0.120 & 4.21 & 0.5 & 0 \\
60318 & 1896.0699 & 1.020 & 324.20 & 0.101 & 3.56 & 0.7 & 0 \\
60318 & 1896.8700 & 1.000 & 329.00 & 0.081 & 2.84 & 1.1 & 0 \\
60318 & 1896.8700 & 1.000 & 329.00 & 0.081 & 2.84 & 1.1 & 0 \\
60318 & 1898.0900 & 0.820 & 328.70 & 0.065 & 2.29 & 1.7 & 0 \\
60318 & 1898.2000 & 0.780 & 331.70 & 0.074 & 2.61 & 1.3 & 0 \\
60318 & 1898.2500 & 0.720 & 331.50 & 0.101 & 3.56 & 0.7 & 0 \\
60318 & 1898.2900 & 0.980 & 329.00 & 0.095 & 3.33 & 0.8 & 0 \\
60318 & 1899.0601 & 0.730 & 333.40 & 0.089 & 3.14 & 0.9 & 0 \\
60318 & 1899.0800 & 0.450 & 330.00 & 0.109 & 3.85 & 0.6 & 0 \\
60318 & 1899.1500 & 0.610 & 334.50 & 0.101 & 3.56 & 0.7 & 0 \\
60318 & 1900.0800 & 0.660 & 330.10 & 0.085 & 2.98 & 1.0 & 0 \\
60318 & 1900.1899 & 0.710 & 329.00 & 0.120 & 4.21 & 0.5 & 0 \\
60318 & 1901.1700 & 0.750 & 338.40 & 0.089 & 3.14 & 0.9 & 0 \\
60318 & 1901.2300 & 0.630 & 332.60 & 0.081 & 2.84 & 1.1 & 0 \\
60318 & 1901.3199 & 0.610 & 324.90 & 0.134 & 4.71 & 0.4 & 0 \\
60318 & 1902.2000 & 0.480 & 325.30 & 0.109 & 3.85 & 0.6 & 0 \\
60318 & 1902.2100 & 0.640 & 325.60 & 0.095 & 3.33 & 0.8 & 0 \\
60318 & 1902.2700 & 0.540 & 325.70 & 0.134 & 4.71 & 0.4 & 0 \\
60318 & 1903.1600 & 0.550 & 330.10 & 0.077 & 2.72 & 1.2 & 0 \\
60318 & 1903.2300 & 0.650 & 331.60 & 0.095 & 3.33 & 0.8 & 0 \\
60318 & 1906.8400 & 0.790 & 331.30 & 0.085 & 2.98 & 1.0 & 0 \\
60318 & 1908.2400 & 0.580 & 328.50 & 0.154 & 5.44 & 0.3 & 0 \\
60318 & 1908.2500 & 0.840 & 331.30 & 0.267 & 9.42 & 0.1 & 0 \\
60318 & 1909.1500 & 0.780 & 329.80 & 0.109 & 3.85 & 0.6 & 0 \\
60318 & 1910.0500 & 1.020 & 332.20 & 0.101 & 3.56 & 0.7 & 0 \\
60318 & 1910.0500 & 1.020 & 332.20 & 0.101 & 3.56 & 0.7 & 0 \\
60318 & 1910.1390 & 0.630 & 322.60 & 0.085 & 2.98 & 1.0 & 0 \\
60318 & 1910.1400 & 0.630 & 326.60 & 0.085 & 2.98 & 1.0 & 0 \\
60318 & 1910.2000 & 0.780 & 328.90 & 0.120 & 4.21 & 0.5 & 0 \\
60318 & 1910.2000 & 0.780 & 328.90 & 0.120 & 4.21 & 0.5 & 0 \\
60318 & 1911.1700 & 0.730 & 332.20 & 0.077 & 2.72 & 1.2 & 0 \\
60318 & 1911.9600 & 0.860 & 331.30 & 0.067 & 2.36 & 1.6 & 0 \\
60318 & 1913.2300 & 0.780 & 331.40 & 0.081 & 2.84 & 1.1 & 0 \\
60318 & 1914.1000 & 0.670 & 331.80 & 0.077 & 2.72 & 1.2 & 0 \\
60318 & 1915.1899 & 0.530 & 331.10 & 0.089 & 3.14 & 0.9 & 0 \\
60318 & 1917.3300 & 0.750 & 328.00 & 0.095 & 3.33 & 0.8 & 0 \\
60318 & 1921.1400 & 0.860 & 330.90 & 0.052 & 1.85 & 2.6 & 0 \\
60318 & 1923.2200 & 0.980 & 329.30 & 0.095 & 3.33 & 0.8 & 0 \\
60318 & 1923.3101 & 0.650 & 330.60 & 0.120 & 4.21 & 0.5 & 0 \\
60318 & 1923.3300 & 0.660 & 322.00 & 0.120 & 4.21 & 0.5 & 0 \\
60318 & 1923.9570 & 0.805 & 332.70 & 0.089 & 3.14 & 0.9 & 0 \\
60318 & 1925.0200 & 0.650 & 329.00 & 0.109 & 3.85 & 0.6 & 0 \\
60318 & 1925.1400 & 0.780 & 329.10 & 0.109 & 3.85 & 0.6 & 0 \\
60318 & 1925.1899 & 0.800 & 328.00 & 0.134 & 4.71 & 0.4 & 0 \\
60318 & 1925.2000 & 0.840 & 328.80 & 0.109 & 3.85 & 0.6 & 0 \\
60318 & 1927.1600 & 0.660 & 330.00 & 0.085 & 2.98 & 1.0 & 0 \\
60318 & 1928.2100 & 0.770 & 327.70 & 0.047 & 1.67 & 3.2 & 0 \\
60318 & 1930.4900 & 0.720 & 332.90 & 0.074 & 2.61 & 1.3 & 0 \\
60318 & 1932.9200 & 0.730 & 329.00 & 0.069 & 2.43 & 1.5 & 0 \\
60318 & 1933.1400 & 0.720 & 328.80 & 0.069 & 2.43 & 1.5 & 0 \\
60318 & 1935.3700 & 0.700 & 327.60 & 0.056 & 1.96 & 2.3 & 0 \\
60318 & 1936.1700 & 0.650 & 328.70 & 0.095 & 3.33 & 0.8 & 0 \\
60318 & 1936.2200 & 0.600 & 332.80 & 0.120 & 4.21 & 0.5 & 0 \\
60318 & 1937.0699 & 0.680 & 328.60 & 0.057 & 2.01 & 2.2 & 0 \\
60318 & 1937.0800 & 0.860 & 328.30 & 0.067 & 2.36 & 1.6 & 0 \\
60318 & 1937.8700 & 0.590 & 329.80 & 0.041 & 1.45 & 4.2 & 0 \\
60318 & 1938.1100 & 0.550 & 328.90 & 0.109 & 3.85 & 0.6 & 0 \\
60318 & 1939.2100 & 0.650 & 328.10 & 0.095 & 3.33 & 0.8 & 0 \\
60318 & 1939.2300 & 0.640 & 328.30 & 0.067 & 2.36 & 1.6 & 0 \\
60318 & 1939.2500 & 0.660 & 327.30 & 0.069 & 2.43 & 1.5 & 0 \\
60318 & 1940.2100 & 0.680 & 327.80 & 0.134 & 4.71 & 0.4 & 0 \\
60318 & 1941.2300 & 0.640 & 329.00 & 0.055 & 1.92 & 2.4 & 0 \\
60318 & 1941.8101 & 0.750 & 330.00 & 0.101 & 3.56 & 0.7 & 0 \\
60318 & 1943.3199 & 0.740 & 330.60 & 0.057 & 2.01 & 2.2 & 0 \\
60318 & 1950.1400 & 0.570 & 328.00 & 0.109 & 3.85 & 0.6 & 0 \\
60318 & 1950.1700 & 0.580 & 326.60 & 0.069 & 2.43 & 1.5 & 0 \\
60318 & 1950.1801 & 0.530 & 327.60 & 0.065 & 2.29 & 1.7 & 0 \\
60318 & 1950.1899 & 0.540 & 328.10 & 0.134 & 4.71 & 0.4 & 0 \\
60318 & 1952.3199 & 0.530 & 327.60 & 0.067 & 2.36 & 1.6 & 0 \\
60318 & 1953.2200 & 0.500 & 328.00 & 0.067 & 2.36 & 1.6 & 0 \\
60318 & 1955.1700 & 0.520 & 327.80 & 0.069 & 2.43 & 1.5 & 0 \\
60318 & 1955.2400 & 0.510 & 326.20 & 0.060 & 2.11 & 2.0 & 0 \\
60318 & 1957.1801 & 0.450 & 326.20 & 0.067 & 2.36 & 1.6 & 0 \\
60318 & 1958.4399 & 0.300 & 328.30 & 0.060 & 2.11 & 2.0 & 0 \\
60318 & 1959.1500 & 0.360 & 324.40 & 0.058 & 2.06 & 2.1 & 0 \\
60318 & 1959.2000 & 0.480 & 333.30 & 0.085 & 2.98 & 1.0 & 0 \\
60318 & 1959.2200 & 0.380 & 328.70 & 0.095 & 3.33 & 0.8 & 0 \\
60318 & 1959.6300 & 0.390 & 330.50 & 0.081 & 2.84 & 1.1 & 0 \\
60318 & 1959.9640 & 0.370 & 330.50 & 0.085 & 2.98 & 1.0 & 0 \\
60318 & 1960.1300 & 0.500 & 333.40 & 0.067 & 2.36 & 1.6 & 0 \\
60318 & 1960.1949 & 0.380 & 330.50 & 0.045 & 1.57 & 3.6 & 0 \\
60318 & 1961.1700 & 0.340 & 323.50 & 0.085 & 2.98 & 1.0 & 0 \\
60318 & 1961.1899 & 0.330 & 329.80 & 0.053 & 1.88 & 2.5 & 0 \\
60318 & 1961.4000 & 0.330 & 326.70 & 0.074 & 2.61 & 1.3 & 0 \\
60318 & 1962.0699 & 0.350 & 335.40 & 0.101 & 3.56 & 0.7 & 0 \\
60318 & 1962.1700 & 0.330 & 329.00 & 0.049 & 1.72 & 3.0 & 0 \\
60318 & 1962.9611 & 0.290 & 328.40 & 0.051 & 1.78 & 2.8 & 0 \\
60318 & 1963.1010 & 0.310 & 328.30 & 0.154 & 5.44 & 0.3 & 0 \\
60318 & 1963.1801 & 0.340 & 329.10 & 0.046 & 1.62 & 3.4 & 0 \\
60318 & 1963.2200 & 0.370 & 326.90 & 0.134 & 4.71 & 0.4 & 0 \\
60318 & 1964.2900 & 0.340 & 323.80 & 0.109 & 3.85 & 0.6 & 0 \\
60318 & 1965.1340 & 0.330 & 329.00 & 0.052 & 1.85 & 2.6 & 0 \\
60318 & 1965.1500 & 0.320 & 315.90 & 0.134 & 4.71 & 0.4 & 0 \\
60318 & 1965.1899 & 0.260 & 333.40 & 0.061 & 2.16 & 1.9 & 0 \\
60318 & 1965.6000 & 0.300 & 329.00 & 0.095 & 3.33 & 0.8 & 0 \\
60318 & 1965.9900 & 0.340 & 331.00 & 0.039 & 1.39 & 4.6 & 0 \\
60318 & 1966.1300 & 0.350 & 333.80 & 0.071 & 2.52 & 1.4 & 0 \\
60318 & 1966.1429 & 0.240 & 324.80 & 0.065 & 2.29 & 1.7 & 0 \\
60318 & 1966.1600 & 0.260 & 325.50 & 0.061 & 2.16 & 1.9 & 0 \\
60318 & 1968.3149 & 0.220 & 308.60 & 0.267 & 9.42 & 0.1 & 0 \\
60318 & 1968.9399 & 0.250 & 339.60 & 0.061 & 2.16 & 1.9 & 1 \\
60318 & 1969.0760 & 0.180 & 325.90 & 0.089 & 3.14 & 0.9 & 0 \\
60318 & 1969.2700 & 0.200 & 302.00 & 0.267 & 9.42 & 0.1 & 0 \\
60318 & 1975.1200 & 0.100 & 258.00 & 0.267 & 9.42 & 0.1 & 0 \\
60318 & 1976.8577 & 0.046 & 157.70 & 0.074 & 2.61 & 1.3 & 0 \\
60318 & 1976.9233 & 0.052 & 157.10 & 0.065 & 2.29 & 1.7 & 0 \\
60318 & 1977.0872 & 0.059 & 156.50 & 0.056 & 1.96 & 2.3 & 0 \\
60318 & 1977.9146 & 0.081 & 150.60 & 0.081 & 2.84 & 1.1 & 0 \\
60318 & 1977.9172 & 0.089 & 151.00 & 0.071 & 2.52 & 1.4 & 0 \\
60318 & 1978.1492 & 0.088 & 154.00 & 0.037 & 1.29 & 5.3 & 0 \\
60318 & 1979.7710 & 0.131 & 152.60 & 0.026 & 0.93 & 10.3 & 0 \\
60318 & 1980.1536 & 0.141 & 150.90 & 0.025 & 0.87 & 11.7 & 0 \\
60318 & 1980.7292 & 0.151 & 150.70 & 0.023 & 0.82 & 13.3 & 0 \\
60318 & 1980.7828 & 0.147 & 149.30 & 0.025 & 0.90 & 11.0 & 0 \\
60318 & 1980.8824 & 0.154 & 150.20 & 0.039 & 1.37 & 4.7 & 0 \\
60318 & 1981.1500 & 0.160 & 145.50 & 0.154 & 5.44 & 0.3 & 0 \\
60318 & 1981.9910 & 0.160 & 146.70 & 0.089 & 3.14 & 0.9 & 0 \\
60318 & 1982.8521 & 0.188 & 151.40 & 0.109 & 3.85 & 0.6 & 0 \\
60318 & 1983.0476 & 0.187 & 149.80 & 0.019 & 0.66 & 20.5 & 0 \\
60318 & 1983.0601 & 0.210 & 149.10 & 0.109 & 3.85 & 0.6 & 0 \\
60318 & 1983.2100 & 0.220 & 145.50 & 0.085 & 2.98 & 1.0 & 0 \\
60318 & 1983.9341 & 0.261 & 157.00 & 0.085 & 2.98 & 1.0 & 0 \\
60318 & 1983.9395 & 0.242 & 159.50 & 0.089 & 3.14 & 0.9 & 1 \\
60318 & 1984.0526 & 0.198 & 150.10 & 0.018 & 0.62 & 23.2 & 0 \\
60318 & 1984.0699 & 0.210 & 146.20 & 0.120 & 4.21 & 0.5 & 0 \\
60318 & 1984.7870 & 0.204 & 150.30 & 0.101 & 3.56 & 0.7 & 0 \\
60318 & 1985.1829 & 0.209 & 149.10 & 0.017 & 0.59 & 25.8 & 0 \\
60318 & 1985.7450 & 0.225 & 143.90 & 0.095 & 3.33 & 0.8 & 0 \\
60318 & 1985.8491 & 0.212 & 149.20 & 0.016 & 0.58 & 26.8 & 0 \\
60318 & 1986.1899 & 0.220 & 144.90 & 0.069 & 2.43 & 1.5 & 0 \\
60318 & 1986.2460 & 0.220 & 149.20 & 0.071 & 2.52 & 1.4 & 0 \\
60318 & 1986.8894 & 0.212 & 148.20 & 0.016 & 0.58 & 26.8 & 0 \\
60318 & 1987.1541 & 0.250 & 146.60 & 0.061 & 2.16 & 1.9 & 0 \\
60318 & 1987.2690 & 0.214 & 148.70 & 0.016 & 0.57 & 27.4 & 0 \\
60318 & 1988.9098 & 0.210 & 148.50 & 0.035 & 1.22 & 6.0 & 0 \\
60318 & 1989.2140 & 0.260 & 142.90 & 0.060 & 2.11 & 2.0 & 0 \\
60318 & 1989.2295 & 0.210 & 148.20 & 0.016 & 0.58 & 26.3 & 0 \\
60318 & 1990.1010 & 0.240 & 143.10 & 0.058 & 2.06 & 2.1 & 0 \\
60318 & 1990.2699 & 0.202 & 147.90 & 0.017 & 0.61 & 24.2 & 0 \\
60318 & 1990.2754 & 0.201 & 148.30 & 0.017 & 0.61 & 23.9 & 0 \\
60318 & 1991.0272 & 0.209 & 147.00 & 0.089 & 3.14 & 0.9 & 0 \\
60318 & 1991.0353 & 0.219 & 148.00 & 0.081 & 2.84 & 1.1 & 0 \\
60318 & 1991.2500 & 0.198 & 150.00 & 0.089 & 3.14 & 0.9 & 0 \\
60318 & 1991.8943 & 0.198 & 147.70 & 0.018 & 0.62 & 23.2 & 0 \\
60318 & 1992.2142 & 0.183 & 152.00 & 0.101 & 3.56 & 0.7 & 0 \\
60318 & 1992.2202 & 0.180 & 152.00 & 0.101 & 3.56 & 0.7 & 0 \\
60318 & 1992.2227 & 0.179 & 152.00 & 0.101 & 3.56 & 0.7 & 0 \\
60318 & 1992.3068 & 0.191 & 147.80 & 0.018 & 0.64 & 21.4 & 0 \\
60318 & 1993.1967 & 0.187 & 147.30 & 0.019 & 0.66 & 20.5 & 0 \\
60318 & 1994.0925 & 0.176 & 147.40 & 0.030 & 1.05 & 8.0 & 0 \\
60318 & 1995.1490 & 0.169 & 147.20 & 0.031 & 1.09 & 7.5 & 0 \\
60318 & 1995.3131 & 0.167 & 147.30 & 0.031 & 1.10 & 7.4 & 0 \\
60318 & 1995.9216 & 0.162 & 146.50 & 0.032 & 1.12 & 7.1 & 0 \\
60318 & 1996.8638 & 0.150 & 146.00 & 0.035 & 1.23 & 5.9 & 0 \\
60318 & 1997.1257 & 0.151 & 146.10 & 0.035 & 1.22 & 6.0 & 0 \\
60318 & 1997.8275 & 0.143 & 145.30 & 0.036 & 1.28 & 5.4 & 0 \\
60318 & 1997.8301 & 0.144 & 145.50 & 0.036 & 1.27 & 5.5 & 0 \\
60318 & 1999.8182 & 0.135 & 147.60 & 0.074 & 2.61 & 1.3 & 0 \\
60318 & 1999.8835 & 0.122 & 144.00 & 0.041 & 1.44 & 4.3 & 0 \\
60318 & 2002.9993 & 0.090 & 148.00 & 0.065 & 2.29 & 1.7 & 0 \\
\enddata
\tablecomments{
Non-PHASES astrometric measurements from the WDS 
are listed with 1-$\sigma$ measurements uncertainties and weights.  
Column 1 is the HD Catalog number of the triple system, 
column 2 is the decimal year of the observation, columns 3 and 4 are the 
separation in arcseconds and position angle in degrees, respectively, 
columns 5 and 6 are 
the 1-$\sigma$ uncertainties in the measured quantities from columns 3 and 4, 
column 7 is the weight assigned to the measurement, and column 8 is 1 if the 
measurement is a $>$3-$\sigma$ outlier and omitted from the fit, 0 otherwise.
}
\end{deluxetable*}

\begin{deluxetable*}{lrrrrrrl}
\tablecolumns{8}
\tablewidth{0pc} 
\tablecaption{Velocities of 63 Gem A \label{tab::rv_63_Gem}}
\tablehead{ 
\colhead{Day} 
& \colhead{${\rm RV_{Aa1}}$}
& \colhead{$\sigma_{RV, Aa1}$}
& \colhead{${\rm RV_{Aa2}}$}
& \colhead{$\sigma_{RV, Aa2}$}
& \colhead{${\rm RV_{Ab}}$}
& \colhead{$\sigma_{RV, Ab}$}
& \colhead{}\\
\colhead{(HMJD)} 
& \colhead{(${\rm km \, s^{-1}}$)}
& \colhead{(${\rm km \, s^{-1}}$)}
& \colhead{(${\rm km \, s^{-1}}$)} 
& \colhead{(${\rm km \, s^{-1}}$)}
& \colhead{(${\rm km \, s^{-1}}$)} 
& \colhead{(${\rm km \, s^{-1}}$)}
& \colhead{Source}}
\startdata
39157.337 & 106.8 & 3.5 & \nodata & \nodata & \nodata & \nodata & Abt \& Levy \\
39160.386 & -0.7 & 2.7 & \nodata & \nodata & \nodata & \nodata & Abt \& Levy \\
39185.296 & -53.0 & 2.9 & \nodata & \nodata & \nodata & \nodata & Abt \& Levy \\
39551.265 & 100.9 & 2.2 & -102.1 & 10.7 & \nodata & \nodata & Abt \& Levy \\
39908.247 & -22.9 & 2.7 & 93.5 & 10.3 & \nodata & \nodata & Abt \& Levy \\
40197.486 & 3.4 & 2.9 & 47.1 & 6.5 & \nodata & \nodata & Abt \& Levy \\
40198.485 & 59.9 & 4.0 & -21.8 & 7.7 & \nodata & \nodata & Abt \& Levy \\
40280.221 & 90.3 & 1.8 & -69.4 & 9.2 & \nodata & \nodata & Abt \& Levy \\
40281.221 & -53.7 & 1.8 & 82.8 & 10.7 & \nodata & \nodata & Abt \& Levy \\
40518.545 & -61.8 & 2.2 & 139.4 & 13.8 & \nodata & \nodata & Abt \& Levy \\
40519.507 & 114.7 & 2.2 & -77.9 & 11.9 & \nodata & \nodata & Abt \& Levy \\
40526.497 & -68.1 & 2.2 & 140.9 & 14.6 & \nodata & \nodata & Abt \& Levy \\
40549.428 & -54.1 & 1.8 & 143.6 & 10.3 & \nodata & \nodata & Abt \& Levy \\
40550.508 & 125.9 & 1.3 & -81.7 & 12.6 & \nodata & \nodata & Abt \& Levy \\
40872.520 & -60.2 & 2.2 & 137.2 & 10.7 & \nodata & \nodata & Abt \& Levy \\
40873.475 & 118.2 & 2.4 & -79.0 & 12.3 & \nodata & \nodata & Abt \& Levy \\
40874.486 & -52.5 & 2.7 & 134.0 & 9.2 & \nodata & \nodata & Abt \& Levy \\
40901.503 & -60.0 & 2.7 & 132.4 & 5.4 & \nodata & \nodata & Abt \& Levy \\
41044.143 & -56.4 & 2.0 & 77.9 & 16.1 & \nodata & \nodata & Abt \& Levy \\
41045.176 & 96.8 & 2.0 & -84.2 & 6.9 & \nodata & \nodata & Abt \& Levy \\
46076.850 & 97.7 & 1.0 & -52.7 & 1.6 & 7.3 & 0.6 & Kitt Peak \\
46130.777 & 126.9 & 1.0 & -89.7 & 1.6 & 9.5 & 0.6 & Kitt Peak \\
46389.026 & -46.9 & 1.0 & \nodata & \nodata & 55.9 & 0.6 & Kitt Peak \\
46389.859 & 95.0 & 1.0 & -94.1 & 1.6 & 55.2 & 0.6 & Kitt Peak \\
46390.944 & -51.0 & 1.0 & \nodata & \nodata & 56.9 & 0.6 & Kitt Peak \\
46530.690 & 115.2 & 1.0 & -88.3 & 1.6 & 27.7 & 0.6 & Kitt Peak \\
46531.714 & -78.4 & 1.0 & 138.4 & 1.6 & 26.3 & 0.6 & Kitt Peak \\
46533.725 & -78.4 & 1.0 & 137.4 & 1.6 & 25.4 & 0.6 & Kitt Peak \\
46534.686 & 119.8 & 1.0 & -96.6 & 1.6 & 25.2 & 0.6 & Kitt Peak \\
46722.041 & 125.5 & 1.0 & -83.6 & 1.6 & 5.9 & 0.6 & Kitt Peak \\
46814.903 & 131.2 & 1.0 & -87.6 & 1.6 & 7.8 & 0.6 & Kitt Peak \\
46816.851 & 129.0 & 1.0 & -90.8 & 1.6 & 6.3 & 0.6 & Kitt Peak \\
46868.805 & 108.6 & 1.0 & -64.8 & 1.6 & 8.5 & 0.6 & Kitt Peak \\
46869.737 & -45.6 & 1.0 & 116.7 & 1.6 & 8.4 & 0.6 & Kitt Peak \\
47097.014 & 116.3 & 1.0 & -101.2 & 1.6 & 38.0 & 0.6 & Kitt Peak \\
47098.850 & 106.9 & 1.0 & -88.9 & 1.6 & 38.0 & 0.6 & Kitt Peak \\
47152.056 & -87.7 & 1.0 & 123.4 & 1.6 & 55.4 & 0.6 & Kitt Peak \\
47153.030 & 105.6 & 1.0 & -106.0 & 1.6 & 56.8 & 0.6 & Kitt Peak \\
47247.736 & 111.4 & 1.0 & -103.7 & 1.6 & 44.7 & 0.6 & Kitt Peak \\
47248.769 & -88.4 & 1.0 & 133.2 & 1.6 & 45.0 & 0.6 & Kitt Peak \\
47308.632 & -75.6 & 1.0 & 140.0 & 1.6 & 22.7 & 0.6 & Kitt Peak \\
47310.635 & -76.6 & 1.0 & 140.1 & 1.6 & 21.8 & 0.6 & Kitt Peak \\
48345.670 & 122.9 & 1.0 & -80.8 & 1.6 & 6.1 & 0.6 & Kitt Peak \\
48346.661 & -63.8 & 1.0 & 138.6 & 1.6 & 6.1 & 0.6 & Kitt Peak \\
48506.023 & 123.8 & 1.0 & -91.3 & 1.6 & 18.6 & 0.6 & Kitt Peak \\
48770.648 & 108.8 & 1.0 & -99.2 & 1.6 & 43.9 & 0.6 & Kitt Peak \\
48774.628 & 115.5 & 1.0 & -104.2 & 1.6 & 41.8 & 0.6 & Kitt Peak \\
48912.880 & -69.9 & 1.0 & 144.1 & 1.6 & 9.6 & 0.6 & Kitt Peak \\
53022.316 & 113.3 & 0.635 & -77.3 & 1.57 & 12.4 & 0.6 & AST \\
53051.228 & 122.2 & 0.635 & -89.2 & 1.57 & 15.1 & 0.6 & AST \\
53298.534 & 110.4 & 0.635 & -109.3 & 1.57 & 55.4 & 0.6 & AST \\
53319.500 & 76.1 & 0.635 & -62.7 & 1.57 & 47.7 & 0.6 & AST \\
53329.506 & 113.3 & 0.635 & -101.8 & 1.57 & 42.5 & 0.6 & AST \\
53351.409 & -49.3 & 0.635 & 100.3 & 1.57 & 33.3 & 0.6 & AST \\
53357.399 & -79.0 & 0.635 & 136.7 & 1.57 & 30.7 & 0.6 & AST \\
53395.390 & 96.3 & 0.635 & -63.8 & 1.57 & 20.0 & 0.6 & AST \\
53442.303 & -53.6 & 0.635 & 120.1 & 1.57 & 13.2 & 0.6 & AST \\
53456.296 & -39.0 & 0.635 & 104.8 & 1.57 & 12.3 & 0.6 & AST \\
53486.192 & 107.6 & 0.635 & -67.9 & 1.57 & 9.3 & 0.6 & AST \\
53502.187 & -46.4 & 0.635 & 114.3 & 1.57 & 9.0 & 0.6 & AST \\
53693.554 & -51.4 & 0.635 & 121.3 & 1.57 & 7.9 & 0.6 & AST \\
53723.561 & 116.0 & 0.635 & -77.2 & 1.57 & 8.8 & 0.6 & AST \\
53742.445 & -6.5 & 0.635 & 71.0 & 1.57 & 9.1 & 0.6 & AST \\
53756.297 & 91.4 & 0.635 & -49.7 & 1.57 & 11.0 & 0.6 & AST \\
53769.265 & -64.0 & 0.635 & 136.3 & 1.57 & 11.1 & 0.6 & AST \\
53786.375 & -62.7 & 0.635 & 133.8 & 1.57 & 12.5 & 0.6 & AST \\
53799.329 & 98.5 & 0.635 & -59.4 & 1.57 & 14.3 & 0.6 & AST \\
53818.278 & 117.1 & 0.635 & -84.8 & 1.57 & 15.5 & 0.6 & AST \\
53849.229 & 118.1 & 0.635 & -89.6 & 1.57 & 19.5 & 0.6 & AST \\
53863.178 & 77.6 & 0.635 & -43.6 & 1.57 & 21.4 & 0.6 & AST \\
53877.160 & -57.3 & 0.635 & 115.2 & 1.57 & 23.3 & 0.6 & AST \\
54100.527 & 115.9 & 0.635 & -102.2 & 1.57 & 37.8 & 0.6 & AST \\
54128.395 & -57.6 & 0.635 & 115.0 & 1.57 & 27.1 & 0.6 & AST \\
54845.366 & -52.6 & 0.635 & 95.3 & 1.57 & 43.7 & 0.6 & AST \\
54847.341 & -62.2 & 0.635 & 107.3 & 1.57 & 43.7 & 0.6 & AST \\
54879.190 & 88.2 & 0.635 & -61.6 & 1.57 & 29.1 & 0.6 & AST \\
54905.329 & -53.9 & 0.635 & 116.2 & 1.57 & 22.5 & 0.6 & AST \\
54920.186 & 114.3 & 0.635 & -81.7 & 1.57 & 19.2 & 0.6 & AST \\
54943.220 & 126.3 & 0.635 & -95.2 & 1.57 & 15.5 & 0.6 & AST \\
54952.182 & -42.1 & 0.635 & 106.3 & 1.57 & 15.1 & 0.6 & AST \\
54965.167 & -34.0 & 0.635 & 98.3 & 1.57 & 12.2 & 0.6 & AST \\
55105.456 & 123.4 & 0.635 & -81.7 & 1.57 & 5.6 & 0.6 & AST \\
55119.408 & 85.4 & 0.635 & -41.9 & 1.57 & 6.7 & 0.6 & AST \\
55146.261 & 125.3 & 0.635 & -83.7 & 1.57 & 5.7 & 0.6 & AST \\
55157.462 & 86.6 & 0.635 & -41.6 & 1.57 & 6.1 & 0.6 & AST \\
55199.314 & -72.7 & 0.635 & 148.7 & 1.57 & 7.7 & 0.6 & AST \\
55278.296 & -40.0 & 0.635 & 106.7 & 1.57 & 11.7 & 0.6 & AST \\
55283.231 & 116.2 & 0.635 & -77.3 & 1.57 & 10.9 & 0.6 & AST \\
55287.196 & 126.2 & 0.635 & -92.3 & 1.57 & 11.4 & 0.6 & AST \\
55292.252 & -58.9 & 0.635 & 129.7 & 1.57 & 11.7 & 0.6 & AST \\
55297.219 & 82.9 & 0.635 & -40.3 & 1.57 & 12.3 & 0.6 & AST \\
55311.236 & -58.6 & 0.635 & 126.4 & 1.57 & 13.0 & 0.6 & AST \\
55321.155 & -70.4 & 0.635 & 141.7 & 1.57 & 14.5 & 0.6 & AST \\
\enddata
\tablecomments{
Radial velocity measurements 
of 63 Gem A.
}
\end{deluxetable*}
\clearpage

\end{document}